\documentclass{PoS}
\usepackage{subfigure}
\usepackage{url}
\usepackage{lineno}

\title{MAGIC as a high-energy $\nu_\tau$ detector: performance study to follow-up IceCube transient events}

\ShortTitle{MAGIC as a high-energy tau-neutrino detector}

\author{M. Mallamaci$^{1}$\thanks{manuela.mallamaci@pd.infn.it} , \speaker{D. Gora$^{2}$}, E. Bernardini$^{1,3}$ for the MAGIC Collaboration\footnote{\texttt{https://magic.mpp.mpg.de/}. For collaboration list see PoS(ICRC2019)1177}\\

{$^1$}Padova University and INFN, I-35131 Padova, Italy \\{$^2$}Institute of Nuclear Physics Polish Academy of Sciences, Radzikowskiego 152, Cracow, Poland\\ {$^3$}Deutsches Elektronen-Synchrotron (DESY), D-15738 Zeuthen, Germany
}
\abstract{MAGIC is a system of two Imaging Atmospheric Cherenkov Telescopes located on the Canary Island of 
La Palma and dedicated to the study of very high energy gamma rays above 30 GeV. MAGIC has recently demonstrated
 its capability as a neutrino detector, by exploiting the Earth-skimming technique. The neutrino-event selection has been
  studied by pointing the telescopes towards the sea a few degrees below the horizon, with no pre-defined source direction. 
  An upper limit to the $\nu_\tau$ flux was set. 
In this work, a follow-up strategy of given source directions is presented. This new approach is characterised by pointing
 to targets passing through the sea window observable with MAGIC and for which a trigger from dedicated neutrino observatories
  is issued with large significance. 
 Above a few tens of PeV, the acceptance of MAGIC to $\nu_\tau$ is considerably large 
 and compelling results can be obtained for transient events of short duration. 
The performance of this new method is tested on a sample of data collected by pointing MAGIC for a few days from March to May 2016 
towards the direction of a multi-PeV neutrino that IceCube detected on June 11th 2014 (reconstructed direction (J2000.0) RA: 110.34$^\circ$
 and Dec.: 11.48$^\circ$). A selection cut is studied in order to discriminate $\tau$-lepton-induced air showers from the background
  of very inclined cosmic-ray-induced air showers. An upper limit on the neutrino flux from the above given neutrino direction is presented.}

\FullConference{36th International Cosmic Ray Conference -ICRC2019-\\
		July 24th - August 1st, 2019\\
		Madison, WI, U.S.A.}

\begin{document}
\section{Introduction}\label{sec:intro}
High-energy and ultra-high energy neutrinos are unique cosmic messengers:  they are 
a probe of hadronic mechanisms at work in cosmic-ray sources and they do not suffer magnetic deflections, pointing to their production site. 
Neutrinos are expected to be produced by astrophysical
accelerators mostly through the charged pion decay. In a low density environment, the flavour ratio at the source is therefore 
($\nu_e$:$\nu_\mu$$\nu_\tau$)$\sim$(1:2:0). 
 Because of neutrino oscillations and according to standard production scenarios, an equal flavour ratio is expected at Earth . \\
The IceCube experiment provided the first strong evidence for the existence of a 
flux of astrophysical neutrinos of energies between about 100 TeV and 1 PeV \cite{icec}.  In addition, its 
measurements are consistent with equal fractions of all flavours \cite{ice2,ice3}. 
However, up to now astrophysical $\nu_\tau$ have not been unambiguously tagged.
 The detection of neutrinos of this flavour would be of fundamental importance from 
 both the astrophysics and particle physics point of view,
 being an additional proof of the cosmic origin of these particles and a 
 clear sign of neutrino oscillations, which can shed light on physics beyond the standard model.\\
 Imaging Air Cherenkov Telescopes (IACTs) are able to detect $\nu_\tau$, by using the so-called \textit{Earth-skimming} method \cite{es}.  
 Up-going neutrinos interacting close to the Earth  surface can produce charged leptons.  
 A $\tau$-lepton arising from a $\nu_\tau$  propagates through the Earth. If it emerges and 
  decays in the atmosphere, it induces an air shower that can be measured through IACTs \cite{IACT_es}.\\
MAGIC is a stereoscopic IACT system focused on the study of the gamma-ray universe between about 50 
GeV up to 50 TeV \cite{MAGIC}.  Even if not designed nor optimised for measuring in Earth-skimming mode, MAGIC set an upper limit 
on the $\nu_\tau$ flux in the PeV-EeV energy range by pointing the telescopes to fixed azimuth and zenith (Az -30$^\circ$, Zd 92.5$^\circ$) \cite{nutau_diffuse}. \\
In this work, we present the analysis of the data collected by pointing MAGIC 
in Earth-skimming mode towards a multi-PeV IceCube event  detected on June 11th 2014 (reconstructed direction (J2000.0) RA: 110.34$^\circ$
 and Dec: 11.48$^\circ$)\cite{atel,list}, called: IceCube-140611A.
In Section \ref{triggered_search}, the strategy for analysing candidate neutrino sources with MAGIC is presented: the cuts for selecting the $\nu_\tau$-signal 
are discussed, and the acceptance and the expected event neutrino rate are calculated. In Section \ref{HET},
we apply our analysis on IceCube-140611A data, setting an upper limit on the flux of $\nu_\tau$ from this direction. The aim of this work is to implement 
a tool that can be used for following-up transient events in Earth-skimming 
mode, in case of alerts issued with a large significance from dedicated neutrino observatories.
\section{Search for Earth-skimming neutrinos from transient events with MAGIC}\label{triggered_search}
\begin{figure}[t]
\begin{center}
 \includegraphics[width=13cm]{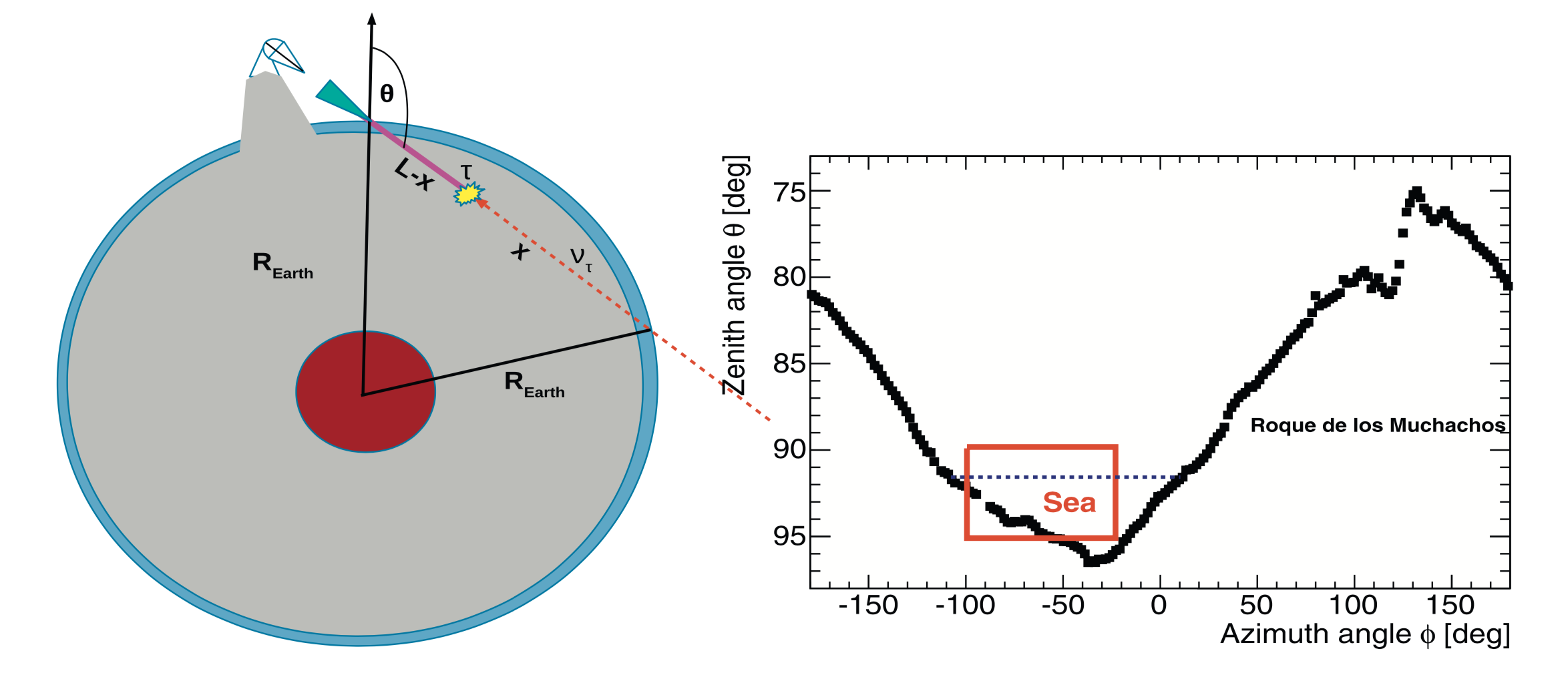}
\caption{\small\rm Left panel: sketch of  $\nu_\tau$-detection in Earth-skimming mode. 
Right panel: horizon seen by the MAGIC telescopes. The rectangle highlights the optimal 
neutrino visibility window. The azimuth range between 0 and -20$^\circ$ is excluded because of the shadowing from
the telescope access tower (see text for more details).
}
\label{window}
\end{center}
\end{figure}

The MAGIC telescopes are able to point down to 6$^\circ$ below the horizontal plane and, thanks to their location, 
have the right distance from the Atlantic Ocean for potentially observing a full development of $\tau$-lepton-induced 
air showers emerging from the sea \cite{IACT_es}. The $\tau$ decay length is about 0.5 (50) km, for an energy of 1 (1000) PeV.  Fig. \ref{window} 
 (left panel) shows a  sketch of an up-going $\tau$-lepton-induced air shower produced by a $\nu_\tau$. The horizon seen 
 by MAGIC is reported in Fig.~\ref{window} (right panel): in particular, the region highlighted by the red rectangle is
  considered optimal for the observations in Earth-skimming mode.
 The zenith and azimuth ranges are respectively [90$^\circ$ - 95$^\circ$] and 
 [-100$^\circ$-20$^\circ$].\\
The effective area of MAGIC for this kind of observations was calculated analytically. For 
an observation angle of 1.5$^\circ$ below the horizon, it ranges from $\sim$10$^3$ m$^2$  at 100 TeV to 6$\times$10$^4$ m$^2$ 
at 300~PeV, reaching 5$\times$10$^5$ m$^2$ at 100 EeV \cite{tau_effA}.\\ 
A search for neutrinos in Earth-skimming mode  has been already carried out with MAGIC. It was performed with about 30 h of data collected by pointing the telescopes below the horizon towards a fixed position (Az -30$^\circ$, Zd 92.5$^\circ$).  A 90\% C.L. upper limit on a hypothetic $\nu_\tau$ point-source flux of  $E^2\Phi(E)$<2.0$\times$10$^{-4}$ GeV cm$^{-2}$ s$^{-1}$ was obtained in the PeV-EeV energy range \cite{nutau_diffuse}. \\
 The predicted emission of neutrinos above PeV from candidate astrophysical sources is quite low (see e.g. \cite{models}).  The MAGIC sensitivity in Earth-skimming mode is limited mainly by the available observation time. However, the perspectives become appealing in case of  short bursting episodes with high luminosities extending to the highest energies, like in case of flaring Active Galactic Nuclei or Gamma-ray Bursts. These objects can exhibit electromagnetic outbursts of duration significantly shorter than months \cite{txs}.
For this reason, pointing MAGIC to potential sources of neutrinos in case of triggers from dedicated observatories like IceCube can significantly improve the probability of detecting $\nu_\tau$. 
\subsection{Monte Carlo simulations}
In order to study the signal expected from an up-going $\nu_\tau$, a full Monte Carlo simulation chain was set up in \cite{IACT_es, nutau_diffuse}, 
based on a modified version of ANIS \cite{gora}, CORSIKA (6.99)\cite{corsika} and the simulation of the response of MAGIC. 
The simulations have been performed  from 1 PeV to 1000 PeV in steps of 0.33 PeV, and for a zenith angle of 87$^\circ$. Even if this analysis is focused
on events with zenith angles larger than 90$^\circ$, this is not a major issue: a previous work demonstrated the universal behaviour of the parameters used for identifying $\tau$-lepton-induced air showers at large zenith angles above 80$^\circ$ \cite{IACT_es, nutau_diffuse}.\\
The expected signature in the MAGIC camera depends on the different decay channels of the $\tau$-lepton, that mainly decays to hadrons, 
pions  and kaons and in 35\% of the cases to electrons and muons.  
In this work, the simulations are used for defining the selection parameters of  $\tau$-lepton-induced air showers,
for studying the identification efficiency (Sec. \ref{opt}) and for evaluating the acceptance of MAGIC (Sec. \ref{acc}). 
 Mainly because of the camera FoV (3.5$^\circ\times$3.5$^\circ$), MAGIC observes a portion of the atmosphere even when it points below the horizon.  The main background
 to a possible neutrino signal is therefore due to very inclined cosmic rays. In the PeV-EeV energy range, these are especially protons. 
 A comparison between $\tau$-lepton-induced air showers and proton-induced air showers has been carried out at MC-level \cite{IACT_es}.
In general, a $\tau$-lepton-induced air shower will contain many more Cherenkov photons with respect to a proton-induced air shower of  
same energy, mainly because its decay vertex is closer to the detector.   
\subsection{MAGIC observations}
\begin{table}
\begin{center}
\begin{tabular}{|c|c|c|}
\hline
{} & \small\rm SeaON  & \small\rm IceCube-140611A\\
\hline
\small\rm Zenith angle & 92.5$^\circ$ & 90$^\circ$-93$^\circ$ \\
\small\rm Azimuth angle & -30$^\circ$  & -80$^\circ$ to -75$^\circ$\\
\small\rm Observation time (h) & 31.5 & 1.34\\
\hline
\end{tabular}
\end{center}
\caption{Summary of data used in this work and collected by using MAGIC in Earth-skimming mode.}
\label{sum}  
\end{table}  
Two main datasets were collected between 2015 and 2016 in Earth-skimming mode, i.e. by pointing MAGIC towards the window highlighted by the rectangle in Fig. \ref{window} (right panel). They are shown in Table \ref{sum}.
The sample called \textit{SeaON} was mainly taken during nights characterised by high cumulus clouds, preventing normal gamma-ray observations.  
SeaON data have been analysed, but no neutrinos have been identified \cite{nutau_diffuse}. 
For this reason, in this work, these data are used as a background sample for the optimisation  of the selection cuts for short-time observations, 
as described in more detail in Sec.~\ref{opt}.   The data named IceCube-140611A were instead collected for a few days in 2016,
 by pointing MAGIC towards the direction of a multi-PeV IceCube  neutrino detected on June 11th 2014 \cite{atel}, and tracking it when it was passing through
  the $\nu_\tau$visibility window.  The effective duration of the observations was 1.34 h.
\subsection{Optimisation of the selection cuts and trigger efficiency study}\label{opt}
\begin{figure}
\centering
{\includegraphics[width=7.3cm]{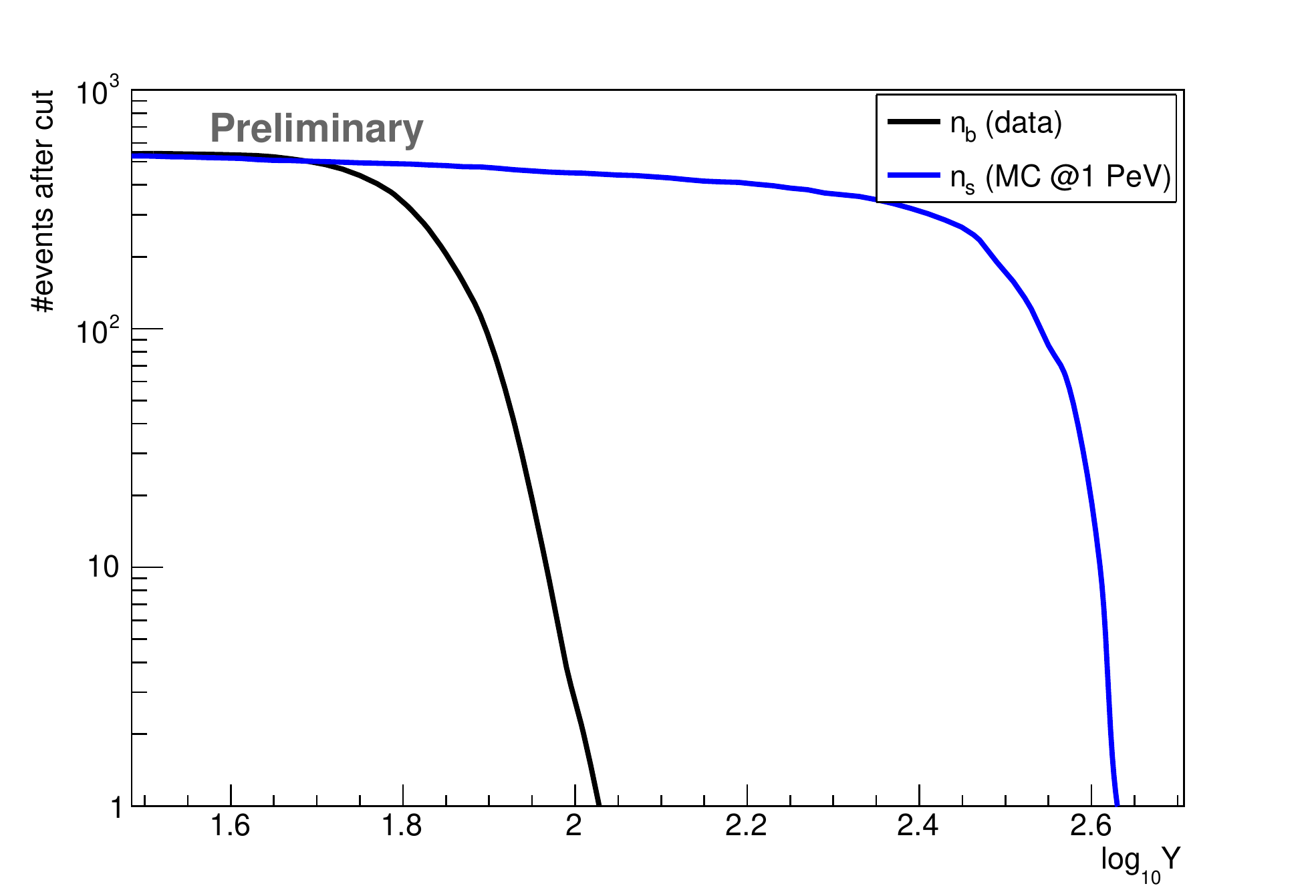}}\quad {\includegraphics[width=7.3cm]{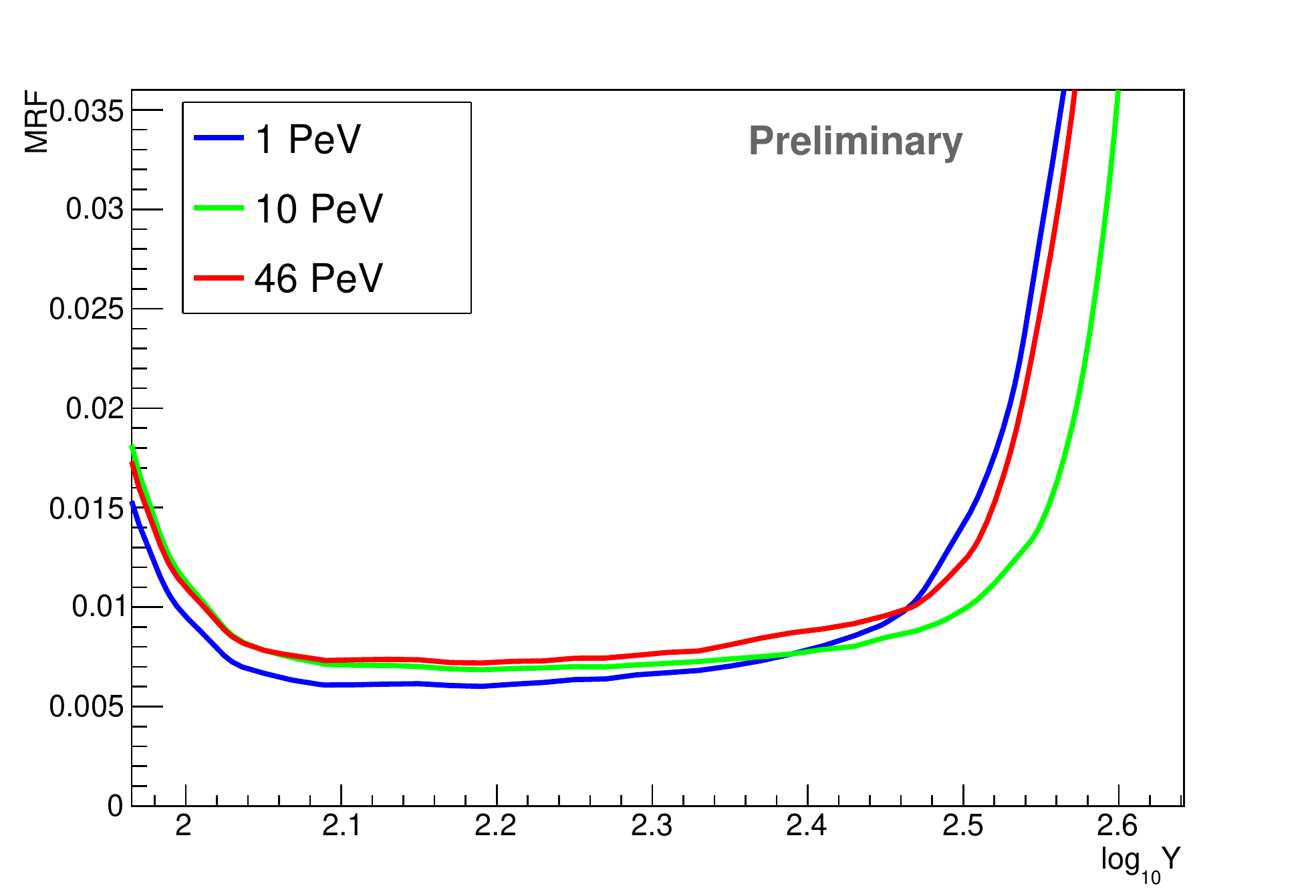}}
\caption{Left panel: cumulative distributions of events as a function of the  log$_{10}$Y cut value. The black line refers
  to background data re-scaled to an equivalent of 1 hour; the blue line represents the MC signal at 1 PeV. Right panel: MRF as a function of log$_{10}$Y. }\label{mrf}
\end{figure}

It is possible to discriminate $\tau$-lepton-induced air showers from the background of very inclined cosmic ray-induced air showers by considering
two Hillas parameters of the cleaned camera image: the \textit{Size} and the \textit{Length} \cite{hillas}.
\begin{figure}
\centering
{\includegraphics[width=7.3cm]{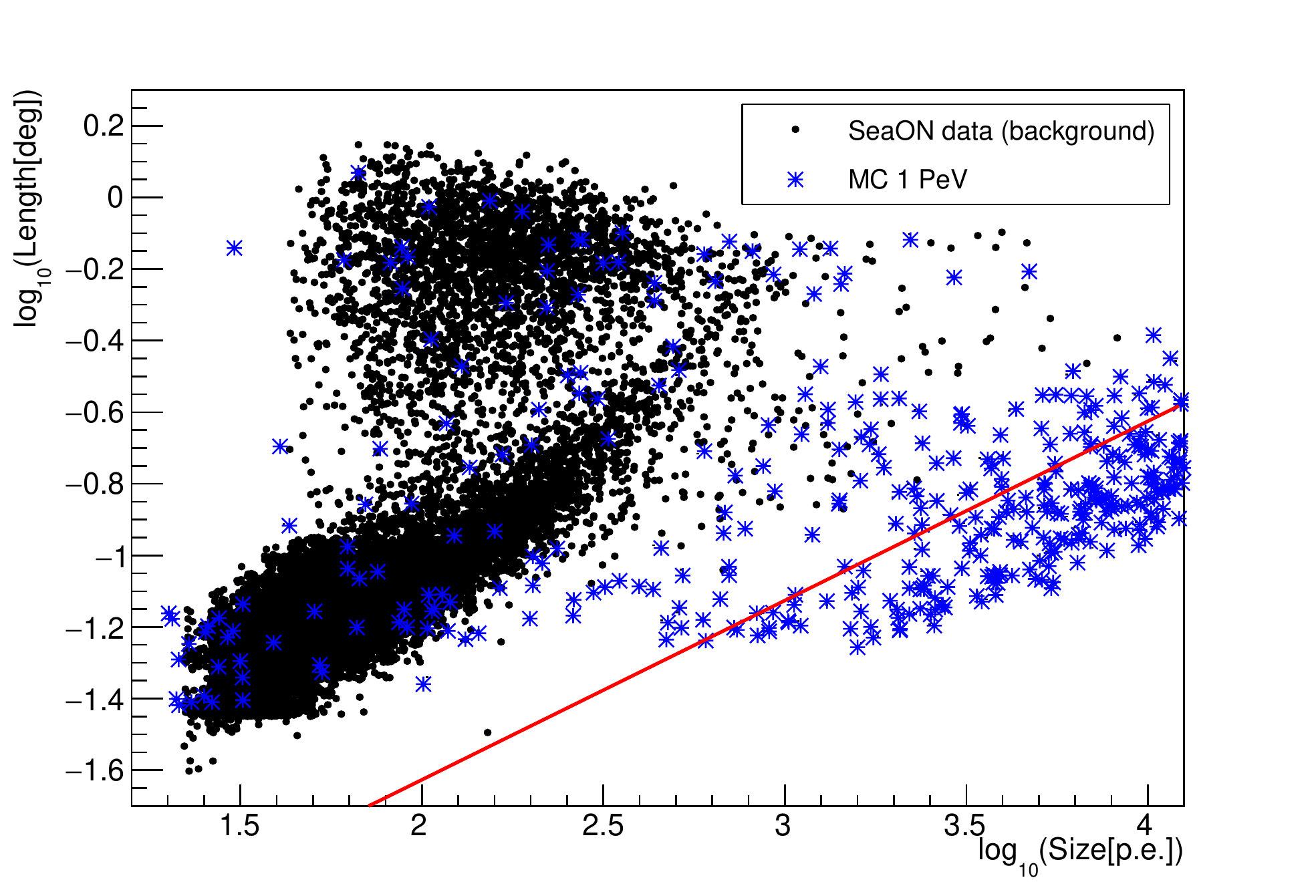}}\quad{\includegraphics[width=7.3cm]{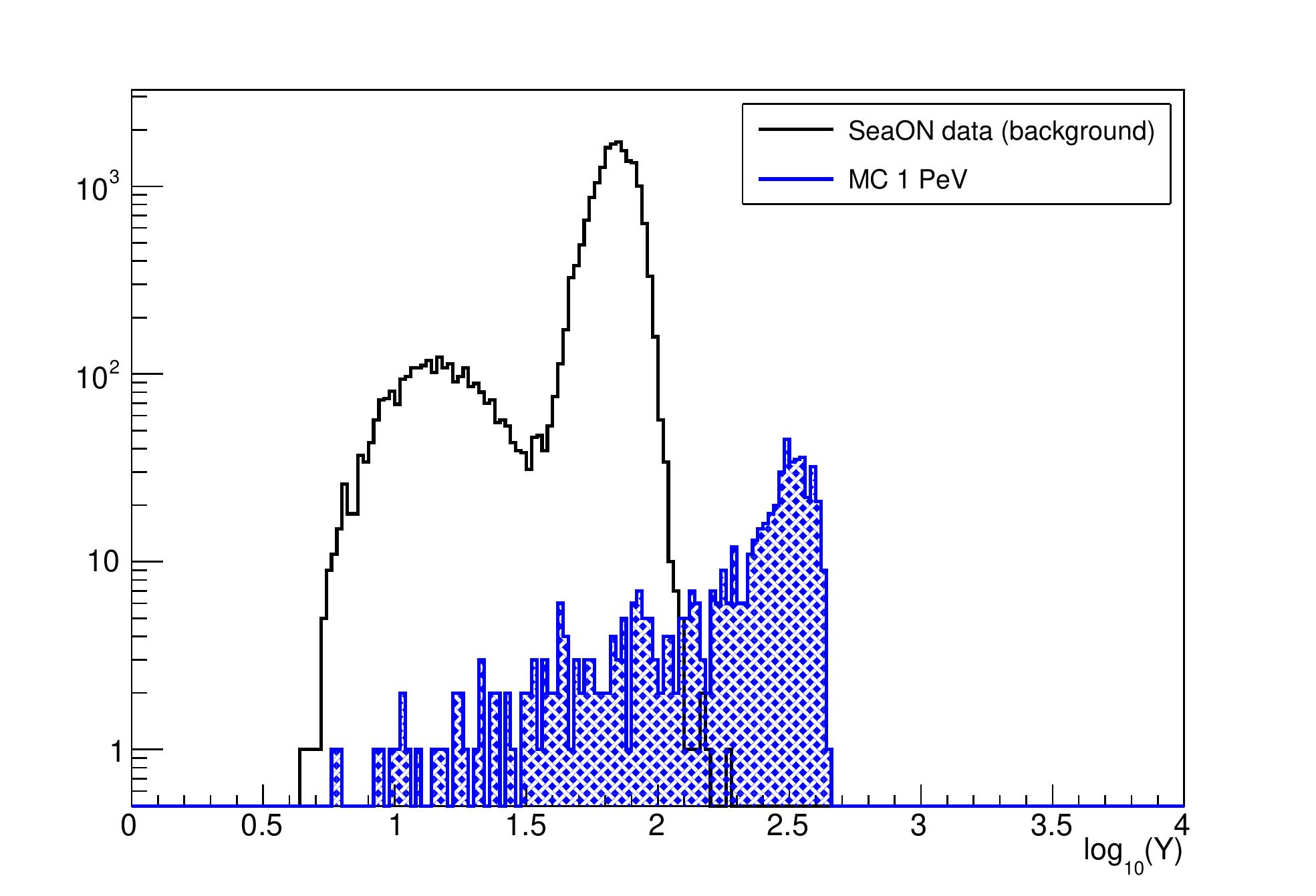}}
 \caption{Left panel: scatter plot of \textit{Size} and \textit{Length} for background data and 
 MC $\tau$-lepton-induced air showers with primary energy of 1 PeV. The red line is the cut adopted in \cite{nutau_diffuse}.
 Right panel: projected distributions in the log$_{10}Y$ direction. See 
 text for details.}\label{logY}
\end{figure}
The first one expresses the total amount of light in the camera, measured in photoelectrons, and it is correlated with the primary energy;
the second quantity describes the lateral development of a shower and it is measured in degrees. 
 Fig. \ref{logY} (left panel) shows a scatter plot of \textit{Size} and \textit{Length}: 
black points are SeaON data, whereas blue points are from MC simulations of $\tau$-lepton-induced air showers at 1 PeV.  Only showers with impact distances 
$d$ < 300 m have been selected, for a better reconstruction of the camera image. 
The figure shows that a region in the 2D \textit{Length}-\textit{Size} plane can be defined that is strongly signal-dominated. The region can be separated by a single cut, by projecting the plane along the direction:
\begin{equation}
\log_{10}Y = \log_{10}(Size)\times\cos\alpha-\log_{10}(Length)\times\sin\alpha
\end{equation}
with $\alpha=63^\circ$. The corresponding projected distributions are shown in Fig. \ref{logY} (right panel).
In \cite{nutau_diffuse}, a cut at log$_{10}Y$=2.35 was set, shown by the red line in Fig. \ref{logY}.
In case of short-duration observations (one-hour time scale),  as those here proposed, 
we expect a smaller background contamination, therefore a reduced cut strength \cite{poster}.  
In particular, the cut has been optimised by studying a \textit{Model Rejection Factor} (MRF) defined as:
\begin{equation}
\rm{MRF}=\frac{\langle\mu^{90}(n_b)\rangle}{n_{s}}
\end{equation}
where $\langle\mu^{90}(n_b)\rangle$ is the average 90\% C.L. upper limit \cite{FC}, $n_b$ is the number of SeaON data
re-scaled to one hour observation time and surviving to log$_{10}$Y cut,  $n_{s}$ is the number of MC $\tau$-lepton-induced 
air showers after the cut.  $n_b$ and $n_{s}$ are shown in Fig. \ref{mrf} (left panel) as a function of the log$_{10}$Y cut. The MRF is shown in Fig. \ref{mrf} (right panel) for different MC energies.  
A method for reconstructing the primary energy of up-going $\tau$-lepton showers has not been implemented yet, however, the figure shows that
 the cut does not depend on it. This is due to the fact that the log$_{10}$Y distributions exhibit a universal behaviour, with a negligible
dependence on energy and also zenith angle \cite{nutau_diffuse}.
The  log$_{10}$Y cut value has been chosen to 
 minimise the ratio between the upper limit setting capability and the expected neutrino signal after the cut.
In particular, we set the cut at log$_{10}Y$=2.1. The cut has been studied also for larger $d$ and we found that it can
 be relaxed, down to log$_{10}Y$=1.9, because of a further reduction of the background.\\ 
The MC signal efficiency is around 80\%, this value is however largely reduced when one considers the trigger efficiency. 
We  estimated it on MC simulations, including the MAGIC response. It is defined as the ratio between the number of simulated showers with positive trigger 
decision and the total number of showers at generation level. It depends on the energy $E_\tau$ of the $\tau$-lepton, 
on the distance $r$ between the $\tau$ decay vertex and the detector, and on the impact distance $d$. The latter has been randomised in CORSIKA and also over the MAGIC camera
FoV \cite{nutau_diffuse}. The total efficiency (trigger+log$_{10}Y$ cut) ranges between 5\% and 20\%, depending on $E_\tau$ and on $r$: in particular, it is larger for larger energies and smaller $r$.
\begin{figure}[h]
 \centering
   {\includegraphics[width=8cm]{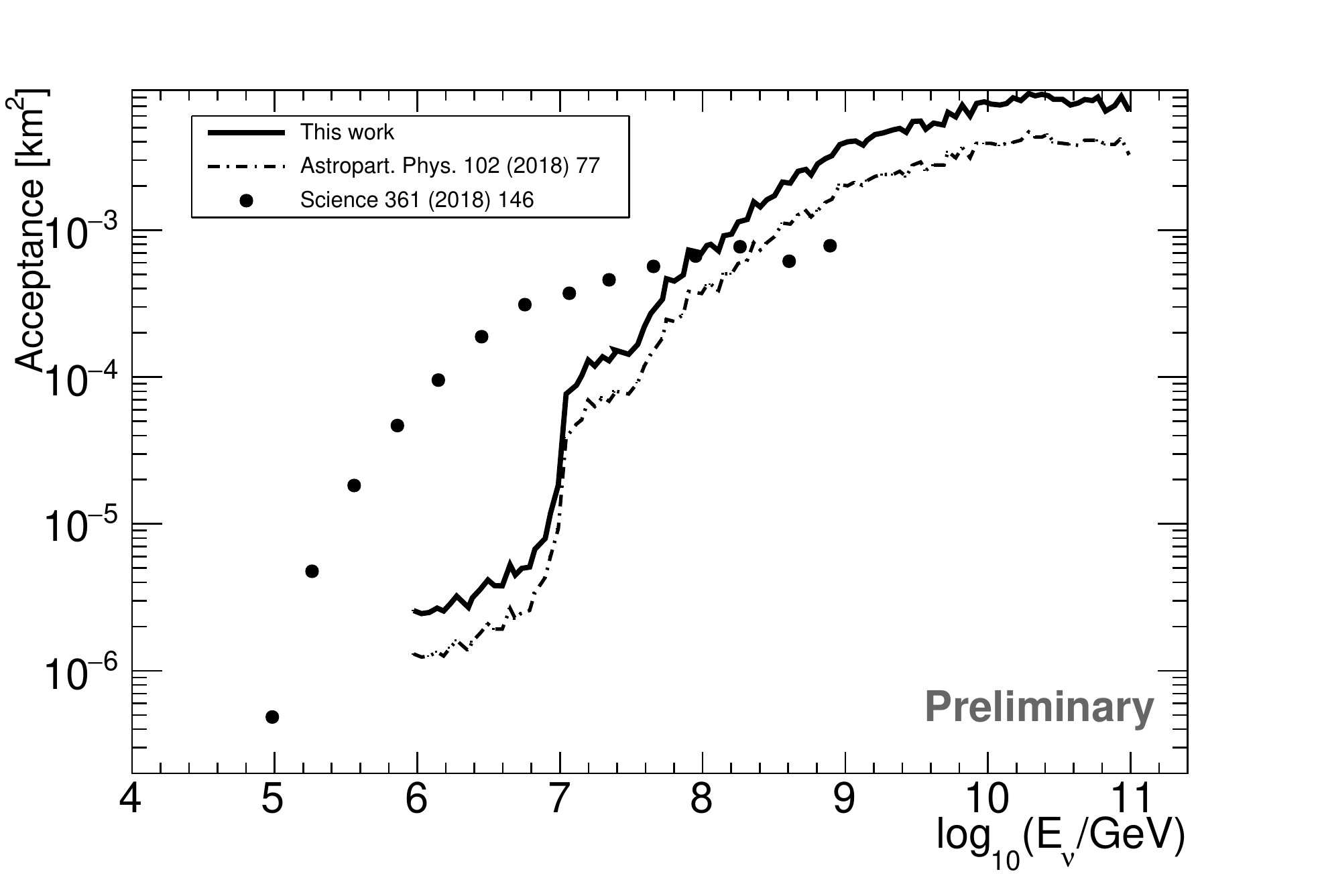}}
 \caption{
 MAGIC acceptance to $\tau$-lepton-induced air showers compared to previous work of MAGIC \cite{nutau_diffuse} and to IceCube acceptance as evaluated in \cite{science}. See text for details. }\label{Acc}
 \end{figure} 
The identification efficiency should depend also on the zenith angle of the shower, but this dependence 
is negligible for $\theta$>83$^\circ$, as demonstrated in~\cite{gora}.
\subsection{Acceptance and event rate calculation}\label{acc}
In order to estimate the neutrino flux from a given source, it is necessary to know the detector acceptance, i.e. the ratio between the number 
of reconstructed events $N_{rec}$ and the number of generated events $N_{gen}$.  
In particular, the point-source acceptance $A^{PS}$ for MAGIC has been evaluated as follows:
\begin{equation}
A^{PS}(E_{\nu_\tau}, \theta,\phi) = \frac{\sum_{i=1}^{N_{FoVcut}}P_i(E_{\nu_\tau},E_\tau, \theta)\times \sigma_i(\theta)\times \epsilon_{\rm{eff,i}}(E_\tau,r,d)}{N_{gen}}.
\end{equation}
$P_i(E_{\nu_\tau},E_\tau, \theta)$ is the probability that a neutrino of
given energy $E_{\nu_\tau}$ and zenith $\theta$ produces a $\tau$-lepton of energy $E_\tau$. $\sigma_i(\theta)$ is the cross-section evaluated for an interaction 
volume given by a cylinder of radius 50 km and height 10 km. The calculation takes into account the density profile of the 
Earth and a water layer of thickness of 3 km (for more details see \cite{gora}).
 $\epsilon_{\rm{eff,i}}(E_\tau,r,d)$ is the trigger+cut efficiency described above. $N_{FoVcut}$ is the number of $\tau$-leptons with decay vertex inside the MAGIC FoV. 
The acceptance of MAGIC is shown in Fig. \ref{Acc}  between 1~PeV and 100~EeV. The dashed line corresponds 
to the acceptance evaluated  with the cuts adopted in \cite{nutau_diffuse}, whereas the solid line is the acceptance evaluated
in this work, showing an improvement by a factor ranging between 1.8 and 2, depending on the energy. The performance of MAGIC is compared to
IceCube. The points in Fig. \ref{Acc} show the
effective area of IceCube evaluated for the online through-going track (''EHE'') selection at the most favourable conditions (in energy and direction) \cite{science}. 
The total number of expected signal events has been finally evaluated by considering the 
flux prediction evaluated for the AGN PKS 2155-304 in high-state activity as a benchmark flux \cite{model_pks}, and the acceptance shown in Fig.~\ref{acc} (solid line). The resulting expected  event rate is $n_s$=1.4$\times10^{-4}/3$h. The systematic uncertainties of the event rate are related to the neutrino-nucleon cross 
section and to the $\tau$-lepton energy loss. They range between 10\% and 40\% as shown in \cite{nutau_diffuse}. 
\section{Analysis of IceCube-140611A with MAGIC}\label{HET}
The analysis described above has been applied on IceCube-140611A observations. About 1.34 hours of data were collected
 by following the source when it was passing in the 
optimal neutrino visibility window. The observations have been performed in standard observation mode, 
usually adopted for gamma-ray
sources. The number of reconstructed stereo events with MAGIC pointing below the horizon (Zd>90$^\circ$) and here selected is 3643. 
\begin{figure}
\centering
 {\includegraphics[width=7.3cm]{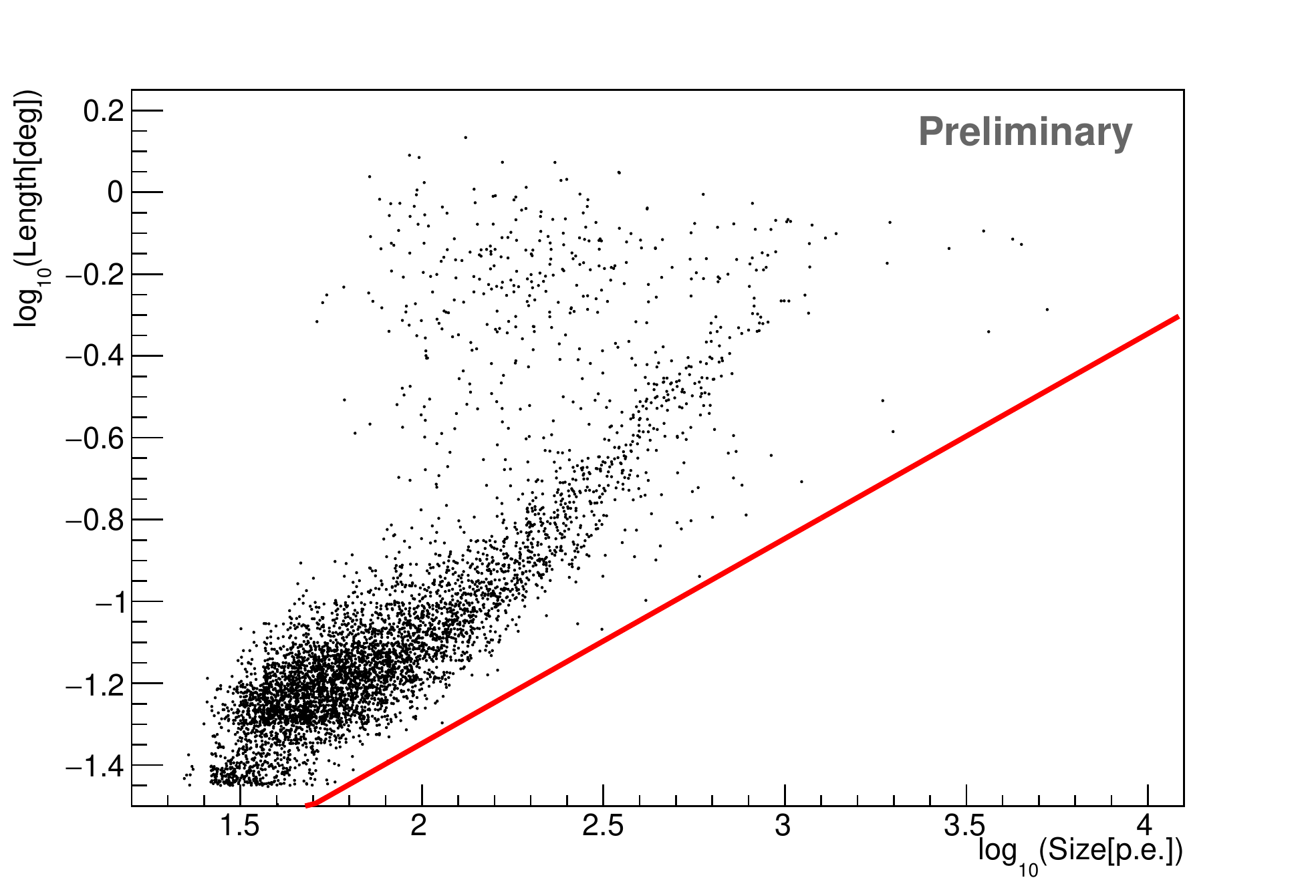}}\quad{\includegraphics[width=7.3cm]{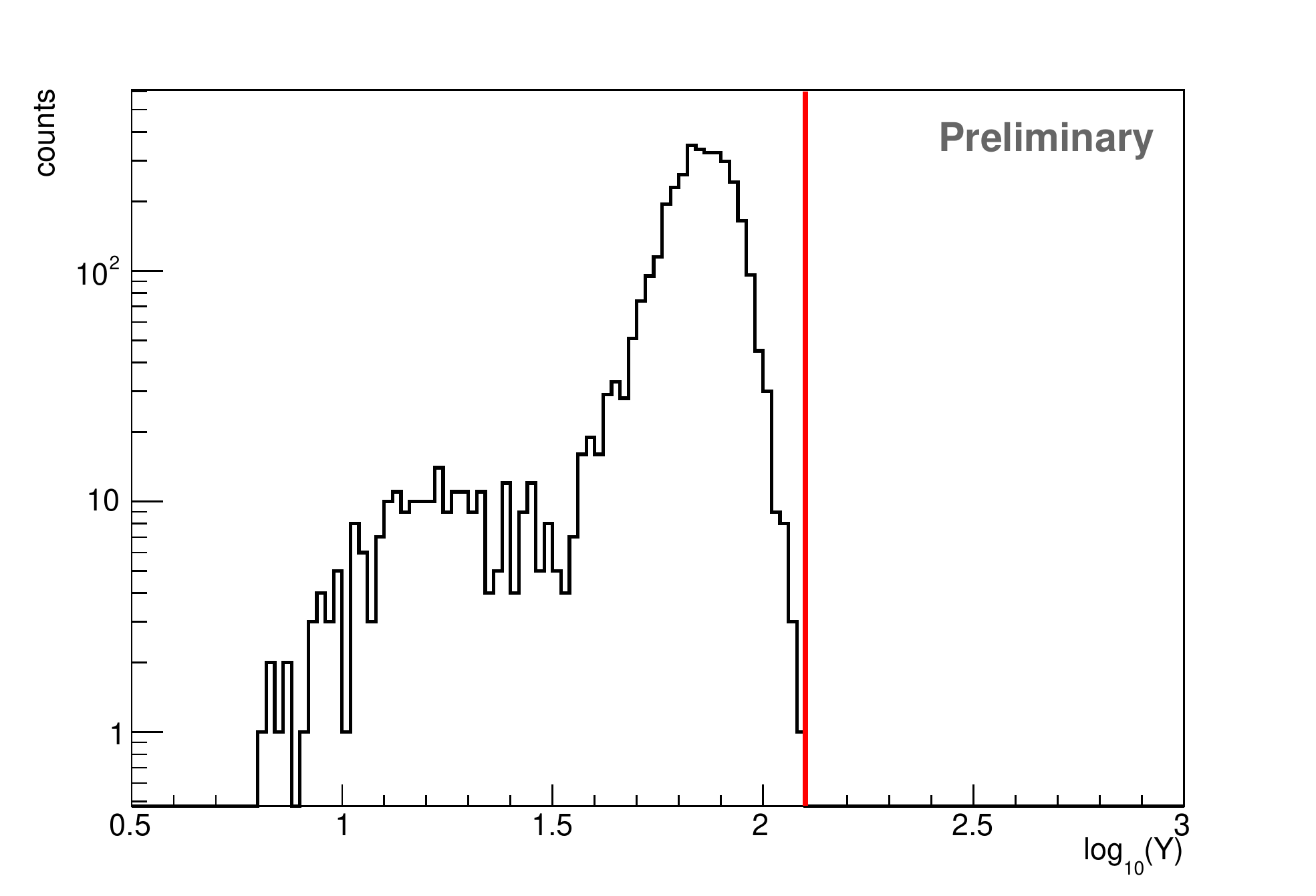}}
\caption{Left panel: scatter plot of \textit{Size} and \textit{Length} for IceCube-140611A data. Right panel: distributions of IceCube-140611A data projected along the selection line log$_{10}$Y. The cut at log$_{10}$Y=2.1 is highlighted by the red line.  
}\label{logYHET}
\end{figure}
The \textit{Size}- and the \textit{Length}-parameters of those data are shown in Fig.~\ref{logYHET} (left panel). 
The corresponding log$_{10}$Y distribution are reported in Fig.~\ref{logYHET} (right panel).\\
The neutrino flux upper limit has been calculated at 90 \% C.L. as follows:
\begin{equation}
\Phi^{\rm{90 \,C.L.}}=\frac{\mu^{\rm{90 \,C.L.}}(n_b, n_{obs})}{n_s}\times\Phi(E_{\nu_\tau}),
\end{equation}
where $\mu^{\rm{90 \,C.L.}}(n_b, n_{obs})$=2.44, as no candidate neutrino events are observed and $\Phi(E_{\nu_\tau})$ is a reference spectrum. 
We assume $\Phi(E_{\nu_\tau})$=10$^{-8}$E$^{-2}$ GeV$^{-1}$cm$^{-2}$s$^{-1}$.
Since  $\Delta$T=1.34 h, the 90\% C.L flux limit on IceCube-140611A is therefore:
\begin{equation}
E^2\Phi^{\rm{90 \,C.L.}}<4\times10^{-4} \rm{GeV \,cm^{-2} s^{-1}}.
\end{equation}
\section{Conclusions}
In this work, we have investigated the capabilities of MAGIC as a high-energy neutrino detector. 
MAGIC can observe neutrinos in the PeV-EeV energy range, through the Earth-skimming channel, by discriminating the  
$\tau$-lepton signal from the background of very inclined cosmic rays. An upper limit on the flux of neutrinos from the direction of IceCube-140611A 
has been calculated.  Even if the sensitivity of MAGIC is limited by the short observation time,  we have demonstrated that for energies above $\simeq$
PeV its acceptance
 is comparable to (or larger than) IceCube effective area (the latter being evaluated in the most favourable conditions \cite{txs}).
For neutrino sources with hard spectra and short outbursts (up to week time scales), MAGIC data-taking in
 Earth-skimming mode can therefore potentially complement the observations by conventional neutrino telescopes, especially
 if the observations are performed following a trigger issued by dedicated experiments like IceCube.
\section*{Acknowledgements}
We would like to thank the agencies and organizations listed here: https://magic.mpp.mpg.de/acknowledgments\_ICRC2019/

\end{document}